\documentclass[11pt]{article}
\usepackage[utf8]{inputenc}
\usepackage[T1]{fontenc}

\usepackage{amsmath, amssymb, amsthm}

\usepackage{graphicx}
\usepackage{caption, subcaption}

\usepackage{natbib}
\usepackage[hidelinks]{hyperref}

\usepackage[a4paper,margin=1.8cm]{geometry}

\title{Fine-Scale Heterogeneity of Snow Cover in the Antarctic Marginal Ice Zone}

\author{
Ippolita Tersigni$^{1}$, Filippo Nelli$^{2{\ddagger}}$, Emiliano Cimoli$^{3}$, Marcello Vichi$^{4,5}$\\
Petra Heil$^{6,1,7}$, Luke G.~Bennetts$^{8}$, Giulio Passerotti$^{1}$, Alberto Alberello$^{9}$\\
Alessandro Toffoli$^{1}$
}

\date{
$^1$Department of Infrastructure Engineering, The University of Melbourne, Parkville, VIC 3010, Australia\\
$^2$Department of Mechanical and Product Design Engineering, School of Engineering, Swinburne University of Technology, Hawthorn, VIC 3122, Australia\\
$^3$Institute for Marine and Antarctic Studies, College of Sciences and Engineering, University of Tasmania, Hobart, TAS 7001, Australia\\
$^4$Department of Oceanography, University of Cape Town, Cape Town, South Africa\\
$^5$Marine and Antarctic Research Centre for Innovation and Sustainability (MARIS), University of Cape Town, Cape Town, South Africa\\
$^6$British Antarctic Survey, High Cross Madingly Rise, Cambridge, CB23 7GB, UK\\
$^7$Australian Antarctic Program Partnership, College of Sciences and Engineering, University of Tasmania, Hobart, TAS 7001, Australia\\
$^8$School of Mathematics and Statistics, The University of Melbourne, Parkville, VIC 3010, Australia\\
$^9$School of Engineering, Mathematics and Physics, University of East Anglia, Norwich NR4 7TJ, UK\\
$[^{\ddagger}$now at Bureau of Meteorology, Science and Innovation Group, Docklands, VIC, 3008, Australia$]$\\[1ex]
}

\begin{document}
\maketitle

\begin{abstract}
Snow cover on Antarctic sea ice is highly heterogeneous, yet climate models have historically simplified it to a uniform layer in winter and bare ice in summer, introducing uncertainty in simulated surface energy fluxes. Here, using shipborne, high-resolution, close-range imagery of the sea ice surface, visual observations, and concurrent atmospheric data from five expeditions across the Antarctic marginal ice zone (2019--2024), we show that fine-scale (sub-grid) snow cover heterogeneity persists across all seasons, is largely independent of ice concentration, and is governed by ice morphology and regional accumulation history. When the heterogeneity is incorporated into flux calculations, it introduces systematic biases in surface energy fluxes relative to uniform-surface assumptions. These findings, based on observations from two Southern Ocean sectors, suggest that the inclusion of sub-grid snow distribution should be considered for simulating Antarctic sea ice evolution and its broader polar climate feedback.
\end{abstract}

\section{Introduction} \label{sec:intro}

Sea ice forms a dynamic, seasonal layer of frozen seawater across the polar oceans, regulating surface energy exchange at the air–sea interface. Its high albedo reflects incoming solar radiation \citep{curry2001applications}, while its insulating properties reduce heat loss from the relatively warm ocean beneath \citep{curry1995seaice, stroeve2012arctic, notz2016observations, tersigni2023high}. Snow cover on sea ice modulates these exchanges, by insulating the ice in winter, thus slowing ice growth \citep{sturm1995snow, maykut1971some}, and reflecting solar radiation in spring and summer, which delays surface warming and melt \citep{wiscombe1980comparison, brandt1993solar, sturm2002winter}.

Snow on Arctic sea ice exhibits relatively structured spatial variability---it is generally thick and uniform on multiyear ice but thinner on seasonal ice, with local variations arising from wind-driven redistribution \citep{Webster2014}. In contrast, Antarctic snow cover is far more heterogeneous because of variable snowfall and stronger wind-driven redistribution \citep{massom2001snow}. 
The heterogeneity  is most severe in the marginal ice zone (MIZ), the outer $\approx{}100$\,km of the ice-covered ocean, where wave-driven fragmentation introduces additional fine-scale heterogeneity at the floe scale \citep{bennetts2022theory,Day2024,egusphere-2025-2184,Fraser2026Preprint}. Floe motions and collisions concentrate snow at floe edges, producing sharp discontinuities, while wave-driven overwash can remove snow entirely from exposed floes \citep{rottier1992floe,pitt2022model,Massom_snow_2025}.

Direct in situ observations of snow on Antarctic sea ice remain scarce \citep{Bourassa_2013,Massom_snow_2025}, limiting understanding of snow distribution and its influence on surface energy fluxes. Numerical models have limited capacity to represent MIZ processes \citep{williams2013wave,horvat2016probability} and, therefore, rely on simplified snow schemes \citep{ecmwf2021ifs,dutra2010snow}. 
For example, the ECMWF Integrated Forecasting System (IFS) has historically assumed sea ice is fully snow-covered in winter and snow-free in summer \citep{ecmwf2021ifs}, introducing uncertainty in simulated albedo and absorbed shortwave radiation \citep{light2008optical,batrak2018modeling,eustace2024evaluating}. Although recent developments have improved the seasonal representation of snow on sea ice \citep[cf.][]{polichtchouk2025ifs50r1}, this floe-scale (or sub-grid) variability in snow cover across different ice types is not yet resolved.

In this Letter, we address this observational gap using a unique multi-season dataset from the Atlantic and Indian sectors of the Southern Ocean, resolving fine-scale heterogeneity in snow and ice surface states across the Antarctic MIZ. Treating the snow–ice surface as a composite system, we quantify how observed snow cover variability modifies radiative and thermodynamic exchanges, identifying key processes that remain insufficiently constrained by observations and commonly simplified in models.

\section{Method} \label{sec:method}

\subsection{Field observations}

\subsubsection{Campaigns and measurement methods} \label{measurement}

\begin{figure}
    \centering
    \includegraphics[width=\textwidth]{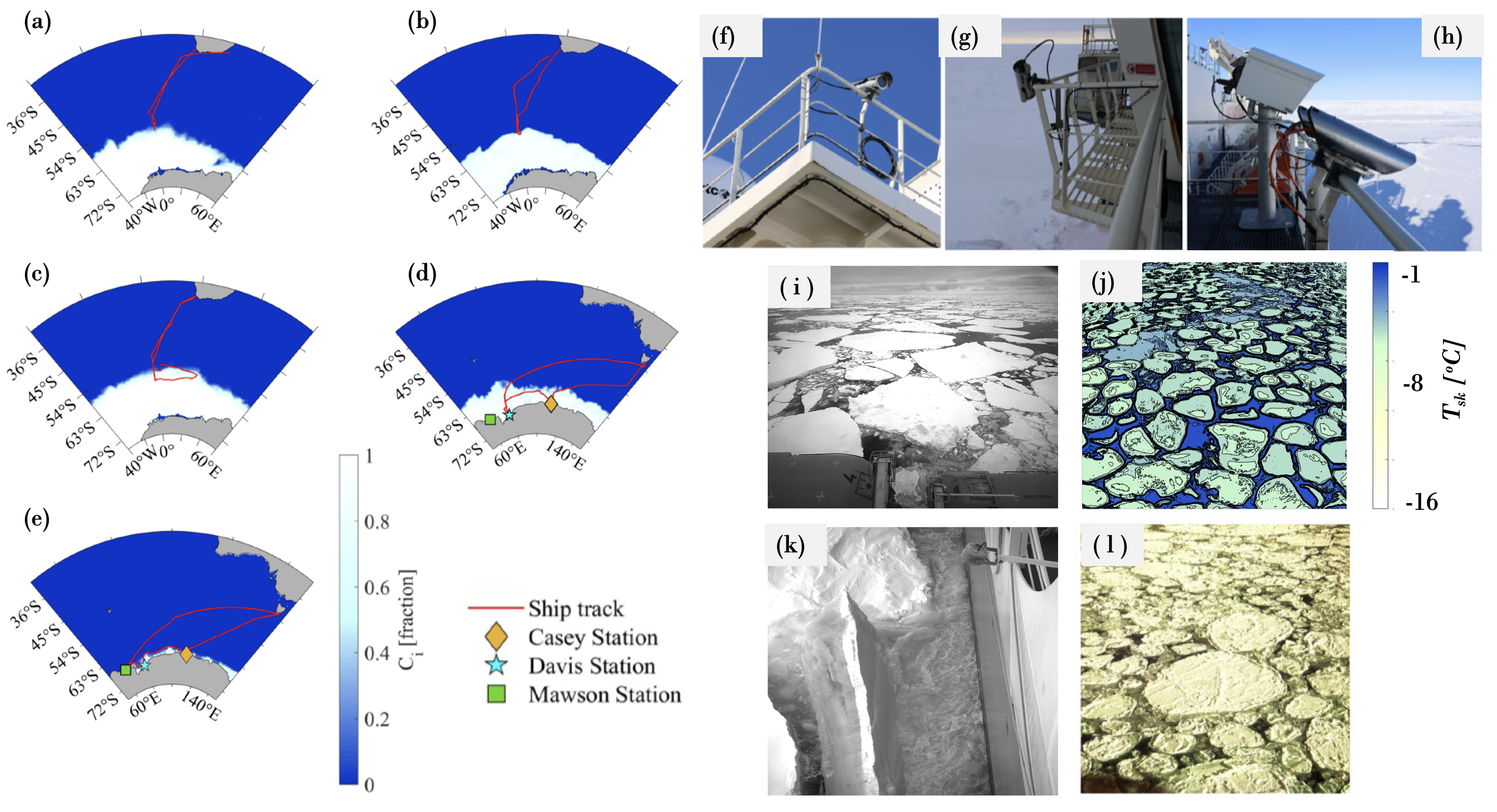}
    \caption{Geographical locations, sea ice concentration ($C_i$; NOAA/NSIDC Near-Real-Time Passive Microwave Climate Data Record), expedition tracks, instrumentation, and sample imagery. (a,b) Winter expeditions, Atlantic sector, \textit{S.A.\ Agulhas II}: (a) July 2019, (b) August 2022. (c,d) Spring expeditions: (c) Atlantic sector, \textit{S.A.\ Agulhas II}, October--November 2019; (d) Indian sector, \textit{RSV Nuyina}, October--November 2024. (e) Summer expedition, Indian sector, \textit{RSV Nuyina}, February--April 2024. (f--h) Shipborne imaging system: (f) wide-angle visible camera, (g) zoom visible camera for ice overturning detection, (h) infrared and multispectral sensors. (i--l) Example imagery: (i) sea ice field, (j) thermal infrared, (k) ice overturning, (l) multispectral reflectance.
    }
    \label{fig:map}
\end{figure}

Observations of sea ice surface properties were acquired during five Antarctic expeditions between 2019 and 2024 aboard two icebreakers \citep{tersigni_2026_20604480}. Three campaigns were carried out in the eastern Weddell Sea (Atlantic sector; Fig.~\ref{fig:map}a--c) aboard the \textit{S.A.\ Agulhas II}, as part of the SCALE project \citep{ryan2022scale,vichi_2023_7902557}, in July 2019 (austral winter), October--November 2019 (spring), and August 2022 (winter). Two further campaigns were conducted aboard \textit{RSV Nuyina} during re-supply voyages to Australian research stations in East Antarctica (Indian sector; Fig.~\ref{fig:map}d,e), in February--April 2024 (summer) and October--November 2024 (spring).

On both vessels, a shipborne imaging suite combining visible and thermal infrared sensors (Fig.~\ref{fig:map}f--h) recorded continuous imagery of the sea ice surface, from which sea ice coverage, floe size, snow-covered fraction, and surface temperature were subsequently derived. The suite comprised a monochrome CMOS visible-light camera \citep{alberello2019brief,alberello2022three,passerotti2026segmenting} and a Telops FAST-IR thermal camera \citep{tersigni2023high}, mounted at 25\,m and 16\,m above the waterline on the \textit{S.A.\ Agulhas II}, and 37\,m and 34\,m on \textit{RSV Nuyina}, both oriented 70$^\circ$ from vertical, pointing toward the sea ice surface. The visible camera recorded 20-minute sequences hourly, whereas the thermal camera recorded 20-minute sequences every three hours (sample images in Fig.~\ref{fig:map}i,j).

On \textit{RSV Nuyina}, additional sensors were deployed. A zoom-lens CMOS camera oriented toward the bow captured overturning floes during icebreaking (Fig.~\ref{fig:map}k) to infer ice thickness and snow depth. A multispectral sensor (400--900\,nm) captured reflected light from the snow cover to characterise the spatial extent of snow-covered ice (Fig.~\ref{fig:map}l).

Imagery was complemented by visual observations of ice type, thickness, and snow depth following the ASPeCt protocol \citep{worby1999,allison1989}---the only method common to all expeditions, and thus the primary in situ source for these variables. Sea ice core thickness was also measured during the \textit{S.A.\ Agulhas II} voyages \citep{ryan2022scale}. Shipboard meteorological systems recorded near-surface air temperature, humidity, wind speed and direction, cloud cover, solar zenith angle, and downwelling PAR (400–700\,nm, inherently accounting for cloud and zenith effects), alongside navigation data and UTC times. All visual, meteorological, and radiative measurements were averaged over 20-minute intervals to match the imagery's temporal resolution.

\subsubsection{Image processing and retrieval of sea ice and snow surface characteristics}

All imagery was georeferenced and quality-controlled to remove frames affected by darkness, poor visibility, lens contamination, or motion blur, retaining 70\% of visible and 64\% of thermal infrared images. Thermal infrared images provide surface temperature at each pixel \citep[instrument validation in][]{tersigni2023high}. Frames were cropped to exclude the ship’s hull, wake, and horizon, and the remaining area was used to estimate mean snow and bare ice temperatures together with an area-weighted surface (skin) temperature \citep{tersigni2023high}.

\begin{figure}
    \centering
    \includegraphics[width=\textwidth]{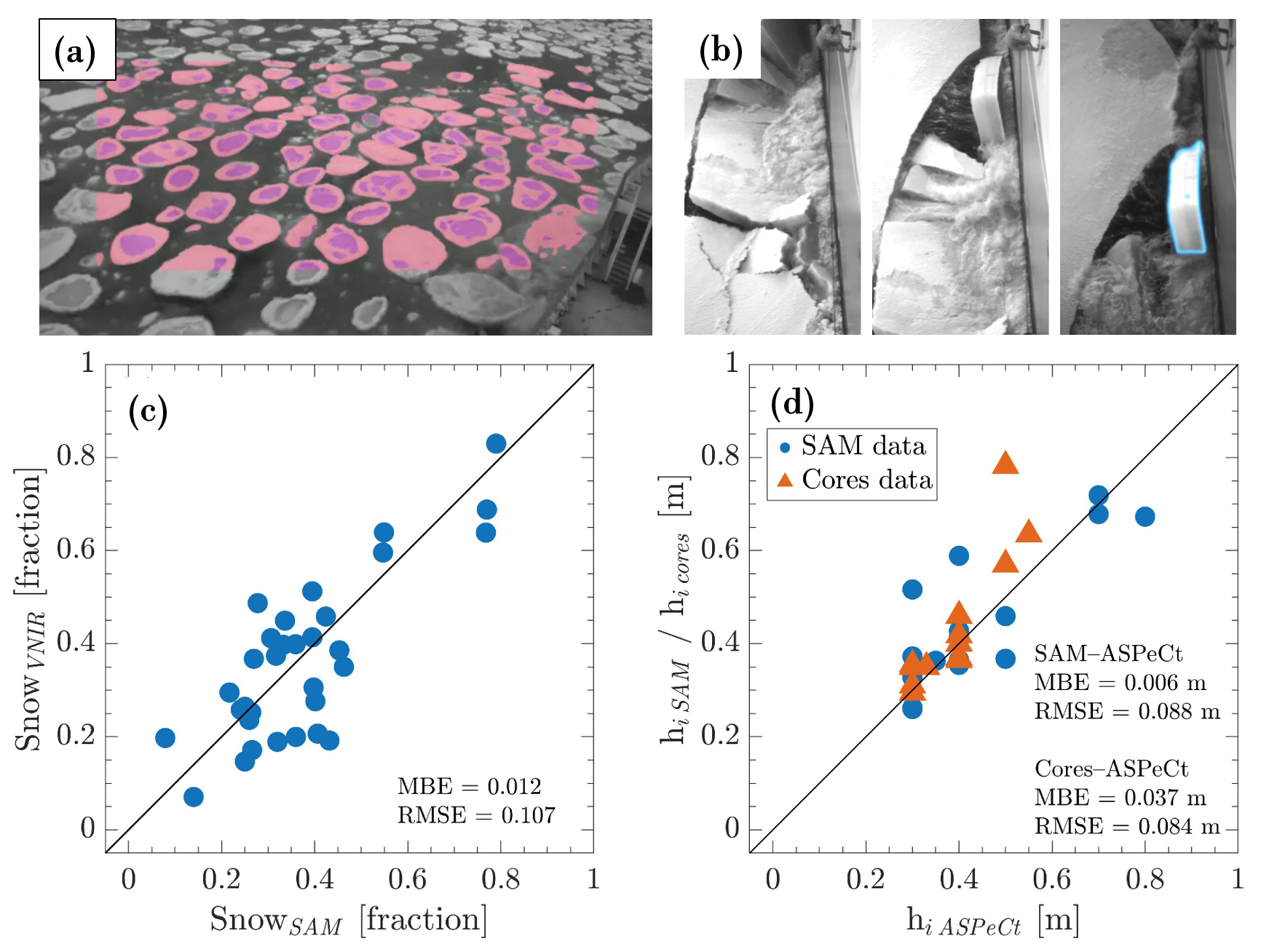}
    \caption{Snow cover and ice thickness ($h_{i}$) evaluation. (a)~Sample image showing identification of snow-covered fractions (light pink) and bare-ice fractions (dark pink) on pancake ice derived using the Segment Anything Model (SAM). (b)~Sequence of overturning ice floes and corresponding segmentation for thickness detection when floes align with the camera axis. (c)~Snow-cover fractions estimated from multispectral imagery compared with values derived from SAM for observations collected during February--April 2024. (d)~Comparison between ASPeCt observations and SAM-derived sea-ice thickness estimates from February--March 2024, together with sea ice core measurements collected in July 2019.
    }
    \label{fig:validation}
\end{figure}

Visible images were orthorectified to correct perspective distortion and provide a consistent ground reference. Orthorectified images were then processed using the Segment Anything Model \citep[SAM;][]{Kirillov2023segment} to isolate ice floes from open water, enabling estimates of sea ice concentration and floe size distributions \citep[see][]{passerotti2026segmenting}. Within segmented floes, an empirically determined grayscale threshold separated snow-covered from bare ice (Fig.~\ref{fig:validation}a), yielding three surface classes in each scene: open water, snow-covered ice, and bare ice.

A separate workflow was applied to a subset of zoom-lens images collected during selected \textit{RSV Nuyina} voyages. SAM was used to identify fully overturned floes and estimate ice thickness from exposed vertical ice profiles (Fig.~\ref{fig:validation}b). These estimates were used solely to validate routine ASPeCt thickness observations (see \S~\ref{val_thickness}) and were not used as primary thickness data.

The accuracy of SAM-based floe segmentation is demonstrated by \citet{passerotti2026segmenting}. Snow cover fractions from visible imagery were validated against synchronized multispectral observations, collected during the February--April 2024 \textit{RSV Nuyina} voyage and derived from reflectance characteristic of snow in the visible and near-infrared spectrum \citep{hall1995development}. The two methods agreed closely (Fig.~\ref{fig:validation}c), with a root mean square error (RMSE) of 10.7\% and a mean bias (MBE) of 1.2\%.

\subsubsection{Validation of visual sea ice thickness estimates}\label{val_thickness}

ASPeCt visual thickness observations involve multiple observers and, therefore, some subjectivity. Validation against overlapping sea ice core measurements (July 2019, \textit{S.A.\ Agulhas II}) and zoom-camera estimates from overturned floes (February–April 2024, \textit{RSV Nuyina}) shows good agreement (Fig.~\ref{fig:validation}d), with RMSE = 0.084\,m and MBE = 0.037\,m for ice cores and RMSE = 0.088\,m and MBE = 0.006\,m for zoom-camera estimates.

\subsection{Estimation of surface albedo and heat fluxes}

\subsubsection{Surface albedo}

Surface albedo ($\alpha$) for dry snow and bare ice was estimated using the parameterisations of Ebert and Curry  \citep[Table 2 of][]{ebert1993intermediate}. For dry snow, albedo combines direct and diffuse radiation components, with the direct component dependent on solar zenith angle, yielding a spectrally integrated albedo of $\approx 0.8$ over the range of solar zenith angles encountered. For bare sea ice, albedo is prescribed as a function of ice thickness, with thinner ice less reflective than thicker ice \citep{brandt2005surface}. Open water albedo is typically low \cite[$<$~0.1;][]{ebert1993intermediate} and is neglected.

The bulk surface albedo, $\alpha_{(\mathrm{Snow\,and\,Ice})}$, was computed as the area-weighted average of dry snow and bare ice albedos based on their fractional coverage within each 20-minute image sequence. Two limiting cases are defined for benchmarking against common numerical model practice: a fully snow-covered surface, $\alpha_{(\mathrm{Dry\,Snow})}$, and a bare ice surface, $\alpha_{(\mathrm{Bare\,Ice})}$, representing growth-season (autumn--winter) and melt-season (spring--summer) conditions, respectively. In this framework, $\alpha_{(\mathrm{Snow\,and\,Ice})}$ reduces to $\alpha_{(\mathrm{Dry\,Snow})}$ under complete snow cover and to $\alpha_{(\mathrm{Bare\,Ice})}$ in its absence.

\subsubsection{Surface energy fluxes}

Following \citet{worby1991ocean}, surface energy fluxes over the ice cover were computed using bulk formulations. For each surface class (dry snow or bare ice), the flux $F_T$ is the sum of radiative, conductive, and turbulent components,
\begin{equation}
F_T = F_{\mathrm{sw}} + F_{\mathrm{lw}} + F_c + F_e + F_s,
\end{equation}
where $F_{\mathrm{sw}}$ and $F_{\mathrm{lw}}$ are net shortwave and longwave radiative fluxes, $F_c$ is conductive heat flux through the snow–ice column, and $F_e$ and $F_s$ are latent and sensible heat fluxes. For a heterogeneous ice cover, the total flux is computed as the area-weighted average of $F_T$ based on the fractional coverage of snow and bare ice.

Net shortwave radiation absorbed at the surface, $F_{\mathrm{sw}}$, follows from incoming shortwave radiation $F_r$ and surface albedo $\alpha$, as
\begin{equation}
F_{\mathrm{sw}} = (1 - \alpha) F_r,
\end{equation}
where $F_r$ is derived from PAR measurements following the conversion of \citet{kirk1994light}.

Net longwave flux, $F_{\mathrm{lw}}$, is the difference between atmospheric longwave input and surface emission, such that
\begin{equation}
F_{\mathrm{lw}} = \epsilon_a \sigma T_a^4 - \epsilon \sigma T_{\mathrm{sk}}^4,
\end{equation}
where $T_a$ and $T_{\mathrm{sk}}$ are air and surface skin temperatures, $\sigma$ is the Stefan–Boltzmann constant, and $\epsilon$ is surface emissivity (0.99 for dry snow and 0.97 for bare ice). Atmospheric emissivity is parameterised as $\epsilon_a = 0.934 + 0.036 \sqrt{e}$, where $e$ is vapour pressure derived from relative humidity and air temperature \citep{murray1967computation,crawford1999improved}.

The conductive flux, $F_c$, follows the layered snow–ice formulation
\begin{equation}
F_c = \frac{k_s}{h_s}(T_s - T_i) + \frac{k_i + \beta S_i/(T_i - 273.15)}{h_i}(T_i - T_b),
\end{equation}
where $k_s$ and $k_i$ are thermal conductivities of snow and ice \citep{maykut1971some,calonne2019thermal,sturm1997thermal}, $h_s$ and $h_i$ are snow depth and ice thickness, $T_s$ and $T_i$ are snow and ice surface temperatures, $T_b = -1.85^\circ$C is the seawater freezing point, $S_i$ is bulk ice salinity, and $\beta$ accounts for salinity dependence. For bare ice, the snow term is omitted.

The latent heat flux, $F_e$, follows the bulk aerodynamic formulation
\begin{equation}
F_e = \rho L_v C_e U (q_s - q_a),
\end{equation}
where $\rho$ is air density, $L_v$ is latent heat of sublimation, $U$ is wind speed at 10\,m, and $q_s$ and $q_a$ are surface and near-surface specific humidity, respectively. The transfer coefficient $C_e$ is 0.0011 over snow and 0.0014 over bare ice \citep{brunke2006intercomparison}.

The sensible heat flux, $F_s$, is defined analogously, as
\begin{equation}
F_s = \rho c_p C_h U (T_{\mathrm{sk}} - T_a),
\end{equation}
where $c_p$ is specific heat capacity of air at constant pressure, $C_h$ is the bulk transfer coefficient, and $T_{\mathrm{sk}}$ is skin temperature \citep{brunke2011assessment}.

\section{Results}

\subsection{Sea ice and snow conditions}

\subsubsection{Winter}

\begin{figure}
    \centering
    \includegraphics[width=\textwidth]{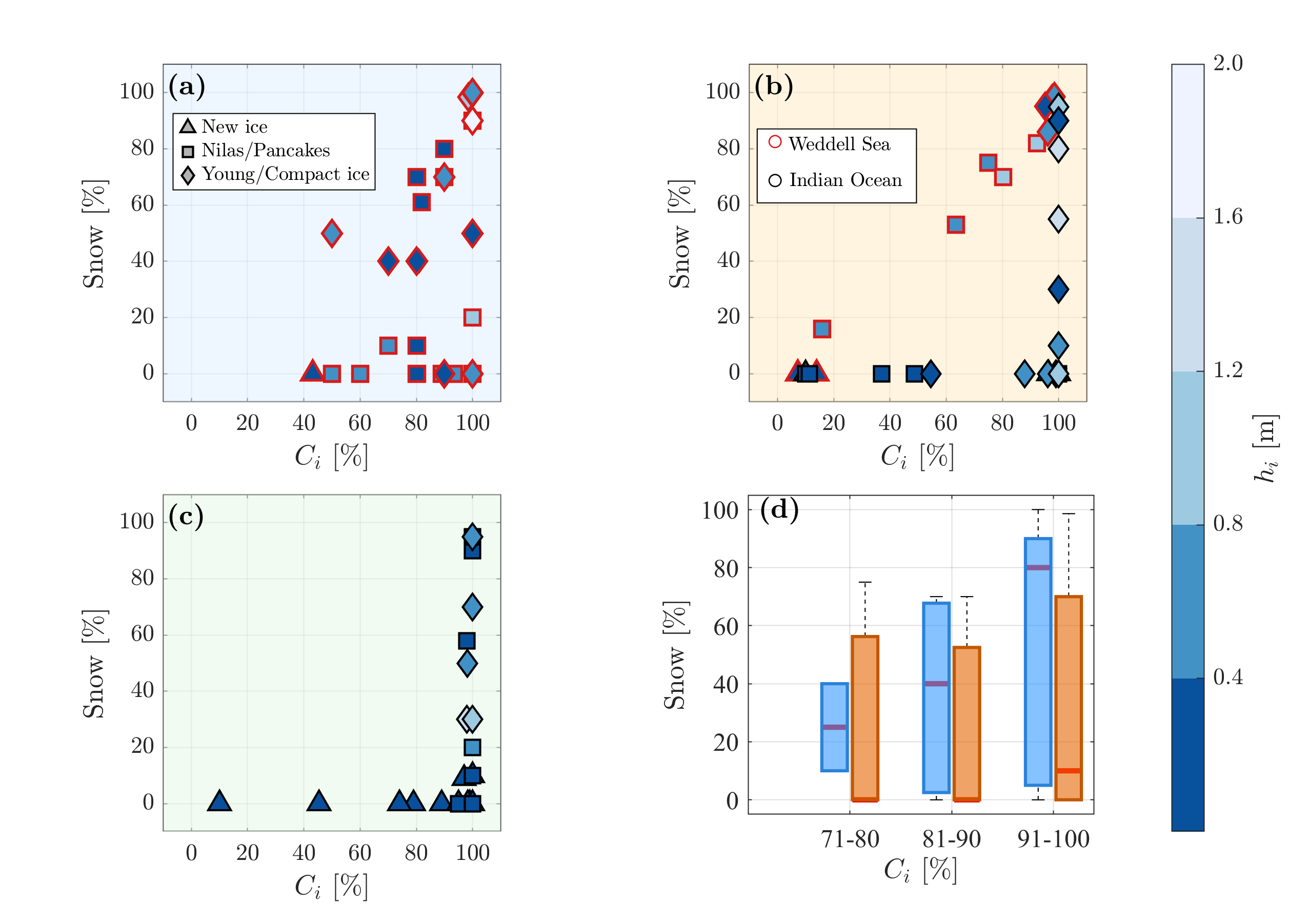}
    \caption{Sea ice concentration ($C_i$) versus fractional snow cover for (a) winter (July 2019 and August 2022, Weddell Sea), (b) spring (October--November 2019, Weddell Sea; October--November 2024, Indian Ocean), and (c) late summer (February--April 2024, Indian Ocean). Sea ice thickness ($h_i$) is colour-coded. (d) Snow cover boxplots by ice concentration class (blue: winter; orange: spring and late summer).
    }
    \label{fig:snow_ic}
\end{figure}

Winter measurements from the eastern Weddell Sea show that snow accumulation reflects the ice's stage of development, from young pancake ice to consolidated pack---floes that have frozen together into a stable, largely continuous cover---rather than a simple dependence on ice concentration (Fig.~\ref{fig:snow_ic}a). As ice concentration increases, snow loading follows two distinct regimes. In one, snow fraction increases with ice concentration, $C_i$, as consolidating floes provide a more stable platform for accumulation \citep{massom2001snow}. In the second, snow cover remains negligible even as $C_i$ approaches 100\%, predominantly in nilas and small pancake ice fields where surface instability, brine wetness, and wave-driven overwash inhibit persistent snow cover \citep{Massom_snow_2025}.

\begin{figure}
    \centering
    \includegraphics[width=\textwidth]{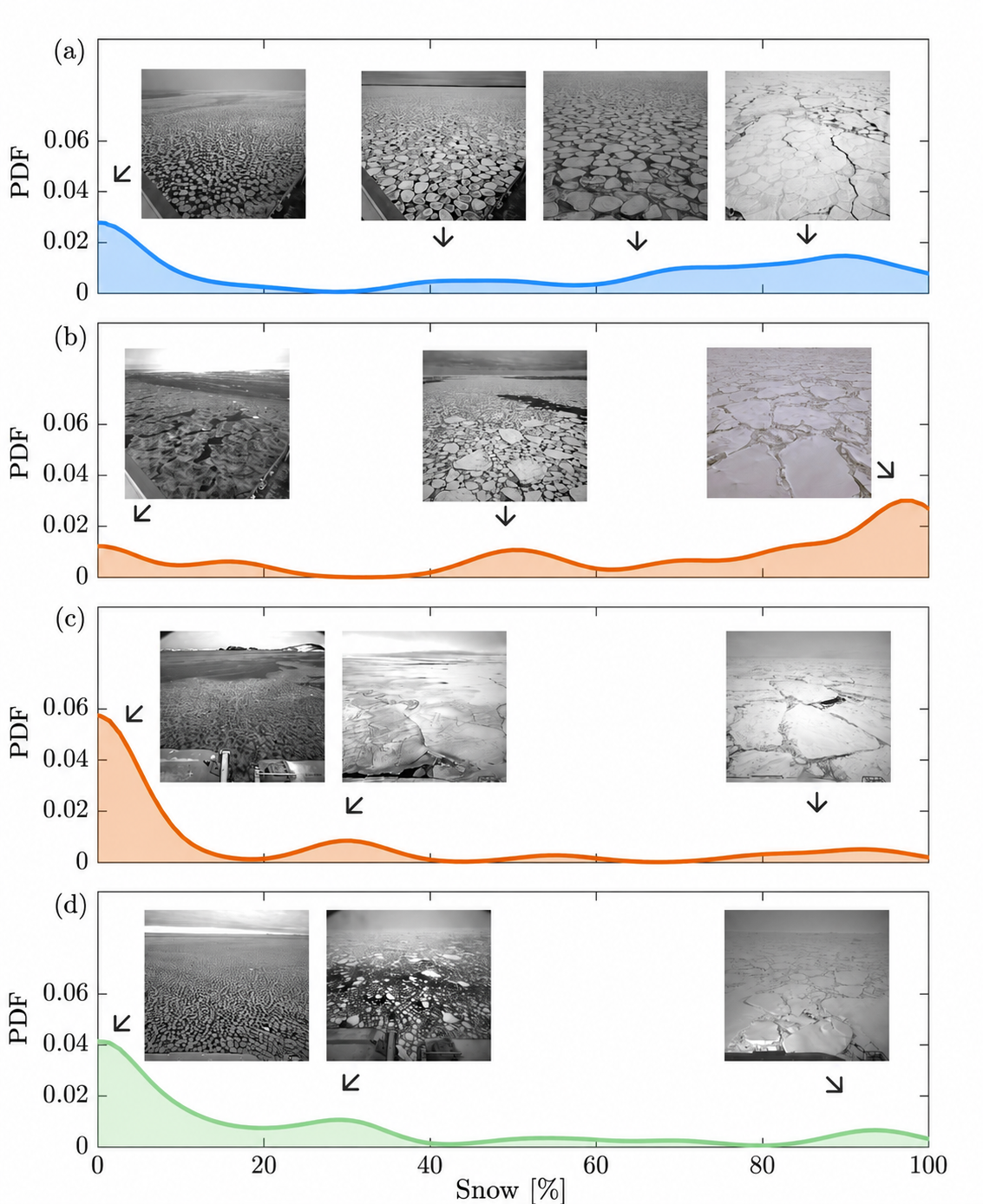}
    \caption{Seasonal distributions of fractional snow cover from visible imagery across five Antarctic expeditions and probability density functions (PDFs) of snow fraction: (a) winter in the eastern Weddell Sea (July 2019 and July 2022); (b) spring in the eastern Weddell Se (October--November 2019); (c) spring in East Antarctica (October--November 2024); and (d) summer in East Antarctica (February--April 2024). 
    }
    \label{fig:snow_distrib}
\end{figure}

Probability density functions (PDFs) of fractional snow cover confirm this partition (Fig.~\ref{fig:snow_distrib}a). The multimodal distributions reflect the heterogeneous ice cover, with a dominant low-snow mode ($<10\%$) representing persistently bare young ice, and intermediate ($\approx$60\%) and high-snow ($\approx$100\%) modes corresponding to increased snow retention on more developed pancakes and consolidated pack ice. This structure indicates that winter snow cover in the MIZ is highly heterogeneous, with coexisting surface states across all ice concentrations.

\subsubsection{Spring}

Spring observations, spanning two different sectors, reveal a clear regional contrast in snow--ice relationships, indicating that seasonal snow loading is not uniform across the Antarctic MIZ (Fig.~\ref{fig:snow_ic}b). In the eastern Weddell Sea, snow cover transitions from the dual winter regimes to a more uniform, near-linear progression, increasing proportionally with ice concentration ($C_i$) and reaching 80–100\% in consolidated ice. This is consistent with precipitation typically peaking in late winter and spring in this sector \citep{arndt2024simulation}. In contrast, the Indian Ocean sector (East Antarctic MIZ) shows a decoupling between ice concentration and snow cover, in which, despite a stable ice platform ($h_i > 0.8$\,m, $C_i > 90\%$), snow distribution remains highly heterogeneous, suggesting that regional factors, such as mesoscale precipitation \citep{massom2015snow} and localized wind scouring \citep{das2015powerful,massom2001snow}, modulate accumulation independently of ice morphology.
These contrasting regimes are reflected in PDFs of fractional snow cover. The eastern Weddell Sea is strongly skewed toward a high-snow mode ($>80\%$; Fig.~\ref{fig:snow_distrib}b), whereas the Indian Ocean is dominated by a 0–10\% peak (Fig.~\ref{fig:snow_distrib}c), indicating that meaningful accumulation is confined to a limited fraction of the consolidated ice cover.

\subsubsection{Late summer}

Observations in the East Antarctic MIZ during late summer indicate a decoupling between ice concentration and snow cover (Fig.~\ref{fig:snow_ic}c), consistent with the snow-free regimes observed during spring in the same region and winter in the eastern Weddell Sea. Despite high ice concentrations ($C_i > 90\%$), snow cover exhibits pronounced spatial heterogeneity, where newly formed pancake ice remains predominantly snow-free, while thicker, older floes ($h_i > 1.0$\,m) span the full range of snow fractions from near-zero to complete coverage. Therefore, ice concentration is  a weak proxy for snow presence in this region. Instead, snow cover is a legacy effect, caused by floes re-incorporated into the pack at different times having accumulated variable snow loads. 

The late-summer PDF (Fig.~\ref{fig:snow_distrib}d) is dominated by a low-snow mode ($<10\%$) associated with advancing pancake fields, with secondary peaks at $\approx$30\% and $\approx$95\% reflecting this variable snow loading on re-incorporated older floes. These results show the snow scarcity identified in spring persists into late summer, maintaining a highly fragmented snow cover not represented by conventional two-state assumptions.

\subsubsection{Seasonal medians and fine-scale variability in snow cover}

The median snow cover (Fig.~\ref{fig:snow_ic}d) follows a clear seasonal cycle across the Antarctic MIZ. During winter, median snow fraction increases from $\approx$25\% to 80\% as the ice field consolidates. During spring and summer, median values remain below 10\% across the full concentration range. These contrasting seasonal medians are broadly consistent with the simplified two-state representation historically used in models \citep{ecmwf2021ifs}, but do not capture the underlying variability of the system. Even at high ice concentration ($C_i > 90\%$), snow cover spans the full range from bare ice to complete coverage within the same concentration class and across seasons. Therefore, seasonal medians  reflect the dominant large-scale cycle while masking the persistent heterogeneity that characterises the Antarctic MIZ at the finer scale.

\subsection{Impact of snow heterogeneity on albedo and heat fluxes}\label{sec:albedo}

\begin{figure}
    \centering
    \includegraphics[width=0.7\textwidth]{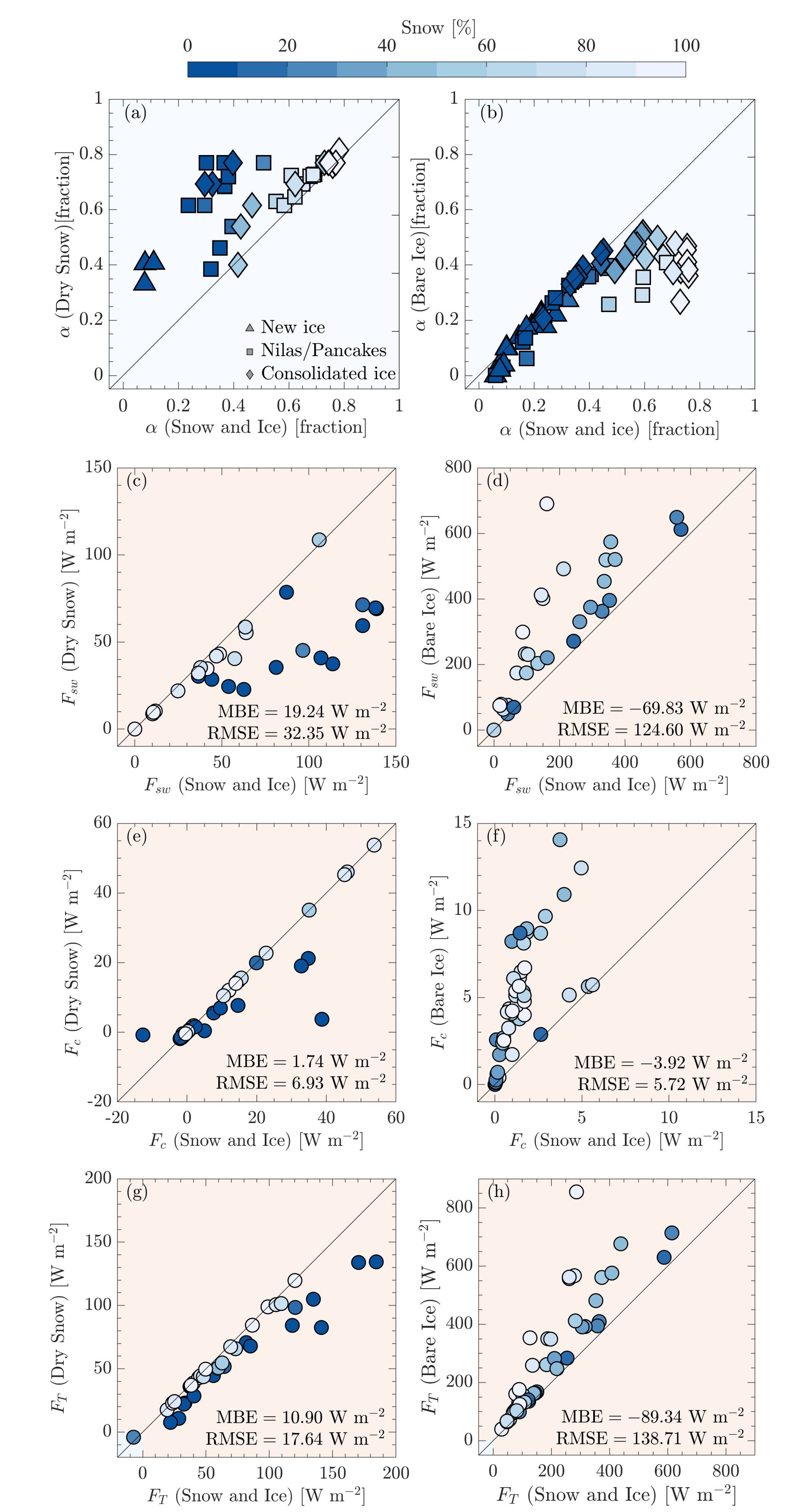}
    \caption{Comparison of surface albedo and energy fluxes between mixed (heterogeneous) and uniform sea ice conditions. (a,b) Effective surface albedo ($\alpha_{\mathrm{Snow\,and\,Ice}}$) relative to dry snow albedo in winter ($\alpha_{\mathrm{Dry\,Snow}}$, a) and bare ice albedo in spring/summer ($\alpha_{\mathrm{Bare\,Ice}}$, b). (c--h) Shortwave ($F_{\mathrm{sw}}$, c,d), conductive ($F_c$, e,f), and total ($F_T$, g,h) surface fluxes for winter (c,e,g) and spring/summer (d,f,h). Colour scale of symbols indicates fractional snow cover throughout (a--h).}
    \label{fig:alb_fluxes}
\end{figure}

Snow heterogeneity produces systematic biases in effective surface albedo and surface energy fluxes relative to uniform snow or bare-ice assumptions, leading to pronounced seasonal asymmetries in surface reflectivity. In winter (Fig.~\ref{fig:alb_fluxes}a), assuming a uniform dry-snow surface ($\alpha_{\mathrm{Dry\ Snow}}$) overestimates effective regional albedo, since the ice cover is often not fully snow-covered. The bias is most pronounced over new ice and young pancake fields, where the observed bare ice patchiness reduces reflectivity by up to 60\% relative to dry snow benchmarks, and persists at $\approx$10\% even as snow cover consolidates. In spring and summer (Fig.~\ref{fig:alb_fluxes}b), the opposite bias occurs, as assuming a bare ice surface ($\alpha_{\mathrm{Bare\ Ice}}$) systematically underestimates effective albedo. Even sparse snow remnants brighten the surface by 10–25\%, reaching $\approx$50\% under substantial snow cover. This persistence indicates that area-averaged schemes mask the sub-grid snow heterogeneity regulating solar absorption during the transition to melt, hence introducing uncertainty in the ice--albedo feedback.

This seasonal asymmetry translates into systematic offsets across the surface energy budget. Figure~\ref{fig:alb_fluxes}c--h shows shortwave ($F_{sw}$), conductive ($F_c$), and total ($F_T$) fluxes, which are most sensitive to snow heterogeneity. Turbulent fluxes are only marginally affected by snow heterogeneity (not shown), since the transfer coefficients for snow-covered and bare ice differ only slightly. The resulting changes in latent and sensible heat exchange remain below 10\%, small relative to the shortwave and conductive contributions.

Shortwave flux shows the strongest response. In winter (Fig.~\ref{fig:alb_fluxes}c), bare ice increases solar absorption relative to a fully snow-covered surface, nearing 100\% in sparsely covered areas. In spring and summer (Fig.~\ref{fig:alb_fluxes}d), even small snow remnants suppress absorption relative to bare ice, and excluding them overestimates solar gain by 50--100\% under high melt-season solar forcing.

Conductive flux shows a similar seasonal divergence, in which deviations are modest in winter (Fig.~\ref{fig:alb_fluxes}e), as snow-free nilas and pancake ice allow rapid ocean–atmosphere exchange, whereas in summer residual snow insulates the ice and suppresses conduction (Fig.~\ref{fig:alb_fluxes}f), such that assuming a uniformly bare surface overestimates conductive flux by nearly 100\%. This insulation also affects the longwave budget via surface temperature, underestimating winter emission and overestimating summer emission by $\approx$20--30\%.

Integrated into total flux, these offsets produce systematic seasonal biases (Figure~\ref{fig:alb_fluxes}g,h). In winter, assuming uniform snow cover underestimates total flux by $\approx$20\% on average, peaking at 50\% where bare ice is present. In spring and summer, the bare ice assumption overestimates net energy gain by $\approx$40\% on average, with local deviations exceeding 60\%. Neglecting sub-grid snow heterogeneity thus leads to an overestimation of summer melt and an underestimation of winter ice growth in surface energy flux estimates across the Antarctic MIZ.

\section{Conclusions}

Analysis of ship-based observations from the eastern Weddell Sea and East Antarctic MIZ, which span winter, spring and late summer, revealed that snow accumulation on ice floes is governed by the ice's stage of development and regional setting, as opposed to a simple area scaling. A transition was observed, from a partitioned winter state to a proportional spring progression in the eastern Weddell Sea, contrasting with persistent decoupling in the East Antarctic MIZ. The results reveal clear regional divergence between the two sectors, particularly in spring, indicating that no single snow–ice relationship applies across the Antarctic MIZ. While seasonal medians reproduce the large-scale cycle represented in simplified two-state model formulations, they mask substantial sub-grid variability, even at near-100\% ice concentration.

This sub-grid variability was shown to introduce systematic, seasonally asymmetric biases in the surface energy budget. Relative to uniform-surface assumptions, observed snow distributions lead to an underestimation of winter heat loss by $\approx$20\% and an overestimation of summer net energy gain by $\approx$40\% on average. These results demonstrate that neglecting sub-grid snow variability leads to biased estimates of seasonal energy exchange, with implications for surface–albedo–energy coupling in models.

Thus, our findings highlight the need to better resolve snow--ice heterogeneity in both observations and models. While consistency across campaigns suggests robustness of the observed patterns, spatially resolved, meter-scale observations linking snow properties to ice type across all Antarctic sectors are lacking. As model representations of snow on sea ice continue to improve \citep{polichtchouk2025ifs50r1}, the observational framework presented here provides a benchmark for evaluating surface processes and supports upcoming satellite altimetry missions targeting polar snow and ice properties.

\section*{Conflict of Interest declaration}
The authors declare no conflicts of interest relevant to this study.

\section*{Open Research}
The data supporting the conclusions of this study are publicly available at \citet{tersigni_2026_20604480} (\url{https://doi.org/10.5281/zenodo.20604480}).

\section*{Acknowledgment}
The expeditions on the \textit{S.A. Agulhas II} were funded by the South African National Antarctic Programme through the National Research Foundation. The financial and logistic support of the South African Polar Research Infrastructure (SAPRI) is acknowledged. Observations from the \textit{RSV Nuyina} were partially supported by Australian Antarctic Science Program grants AAS4434, AAS4506, and AAS4651. Additional financial support was provided by the Australian Research Council through projects LE220100103, DP200102828, FT190100404 and DP240100325. MV was supported by the NRF SANAP contract UID118745. We are indebted to Captain Knowledge Bengu and the crew of the \textit{S.A. Agulhas II} for their invaluable contribution to data collection, as well as the Australian Antarctic Division's technical support group, and captains and crew of the \textit{RSV Nuyina}. All authors acknowledge Dr L. Fascette for technical support. IT thanks H. Bahar, H. Burgess, and I. Law for their contributions to the image analysis in Figure~\ref{fig:validation}a. This research was initiated when PH was at the Australian Antarctic Division (Department of Climate Change, Energy, the Environment and Water, Kingston, TAS, Australia).

\vspace{6pt}
\clearpage


\end{document}